\documentclass[conference,10pt,a4paper]{IEEEtran}

\usepackage{silence}
\newlength{\flexwidth}
\usepackage{bm}
\usepackage{amsmath,amssymb,amsthm,fixmath}
\usepackage{mathtools}
\usepackage{accents}
\usepackage{color,colortbl}
\usepackage{xcolor}
\usepackage{siunitx}
\usepackage{cuted}
\usepackage{multirow,booktabs}
\usepackage{subcaption}
\usepackage[inline]{enumitem}
\usepackage{optidef}
\usepackage{graphicx}
\usepackage{lipsum}

\usepackage{amsmath}
\usepackage{epstopdf}
\usepackage[normalem]{ulem}

\usepackage[textsize=tiny,colorinlistoftodos]{todonotes}
\makeatletter
\define@key{todonotes}{bh}[]{% Comment by 
	\setkeys{todonotes}{author=\textbf{Bin}, color=lime!30}}%
\define@key{todonotes}{zf}[]{% Comment by 
	\setkeys{todonotes}{author=\textbf{Zexin}, color=blue!30}}%
\makeatother

\newif\ifreviewmode
\reviewmodetrue % Set to true to enable review mode
\reviewmodefalse

\ifreviewmode
\else
  \renewcommand{\todo}[1]{} % hide todo notes
\usepackage[shortcuts,acronym,automake]{glossaries}
\makeglossaries
\newacronym{ue}{UE}{User Equipment}
\newacronym{bs}{BS}{base station}
\newacronym{csi}{CSI}{Channel state information}
\newacronym{b5g}{B5G}{Beyond-Fifth-Generation}
\newacronym{6g}{6G}{Sixth Generation}
\newacronym{ml}{ML}{Machine learning}
\newacronym{sbs}{SBS}{small base station}
\newacronym{mu}{MU}{mobile user}
\newacronym{mbs}{MBS}{macro base station}
\newacronym{mse}{MSE}{Mean Squared Error}
\newacronym{cl}{CL}{centralized learning}
\newacronym{uav}{UAV}{Uncrewed Aerial Vehicle}
\newacronym{bme}{BME}{Bayesian Model Ensemble}
\newacronym{iid}{IID}{independent and identically distributed}
\newacronym{raf}{RAF}{robust aggregation function}
\newacronym{sgd}{SGD}{stochastic gradient descend}
\newacronym{cdf}{CDF}{cumulative distribution function}
\newacronym{lid}{LID}{local intrinsic dimensionality}
\newacronym{llpf}{LLPF}{local loss pre-filtering}
\newacronym{mitm}{MITM}{man-in-the-middle}
\newacronym{ae}{AE}{adversary entitie}
\newacronym{tof}{TOF}{time of fly}
\newacronym{rssi}{RSS}{received signal strength}
\newacronym{3d}{3D}{three dimensional}
\newacronym{aoa}{DoA}{Direction of Arrival}
\newacronym{sdp}{SDP}{semi-definite programming}
\newacronym{nlos}{NLOS}{Non-Line-of-Sight}
\newacronym{snr}{SNR}{Signal to Noise Ratio}
\newacronym{crb}{CRB}{Cramer-Rao bound}
\newacronym{lse}{LSE}{least squared estimation}
\newacronym{wlse}{WLSE}{weighted least squared estimation}
\newacronym{gd}{GD}{Gradient descend}
\newacronym{ap}{AP}{Access Points}
\newacronym{crlb}{CRLB}{Cramér-Rao Lower Bound}
\newacronym{tdoa}{TDoA}{Time Difference of Arrival}
\newacronym{sinr}{SINR}{Signal to Interference and Noise Ratio}
\newacronym{los}{LOS}{Line of Sight}
\newacronym{a2g}{A2G}{Air to Ground}
\newacronym{eu}{EU}{European Union}
\newacronym{umiav}{UMi-AV}{Urban Micro–Aerial Vehicle}
\newacronym{3gpp}{3GPP}{3rd Generation Partnership Project}
\newacronym{lae}{LAE}{Low Altitude Economy}
\newacronym{gnss}{GNSS}{Global Navigation Satellite System}
\newacronym{rf}{RF}{Radio Frequency}
\newacronym{gpdr}{GDPR}{General Data Protection Regulation}
\newacronym{5gnr}{5G-NR}{Fifth Generation New Radio}
\newacronym{otdoa}{OTDOA}{Observed Time Difference of Arrival}
\newacronym{prs}{PRS}{Positioning Reference Signals}
\newacronym{gnb}{gNB}{Next Generation Node B}
\newacronym{lmf}{LMF}{Localization Management Function}
\newacronym{amf}{AMF}{Access and Mobility Management Function}
\newacronym{lpp}{LPP}{LTE Positioning Protocol}
\newacronym{nrppa}{NRPPa}{NG-RAN Positioning Protocol A}
\newacronym{prc}{PRC}{Positioning Reference Configuration}
\newacronym{dlotdoa}{DL-OTDOA}{Downlink Observed Time Difference of Arrival}
\newacronym{ulotdoa}{UL-OTDOA}{Uplink Observed Time Difference of Arrival}
\newacronym{nas}{NAS}{Non-Access Stratum}
\newacronym{ngc}{NG-C}{Next Generation Control Plane}
\newacronym{tls}{TLS}{Transport Layer Security}
\newacronym{rsrp}{RSRP}{Reference Signal Received Power}
\newacronym{rof}{ROF}{RSS-based optimum finder}
\newacronym{tcv}{TCV}{Triangular Consistency Verification}
\newacronym{sdet}{SDET}{Static Distance-Error Thresholding}
\newacronym{rdef}{RDEF}{Recursive Distance-Error Filtering}
\newacronym{lawn}{LAWN}{Low-altitude Wireless Network}
\newacronym{ls}{LS}{Least Squares}
\newacronym{iot}{IoT}{Internet of Things}
\newacronym{iq}{IQIA-Net}{In-phase Quadrature Intra-attention Network}
\newacronym{qr}{QR}{Quality report}
\newacronym{ages}{AGES}{Adaptive Gain Exponential Smoother}
\newacronym{v2x}{V2X}{Vehicle-to-Everything}
\newacronym{isac}{ISAC}{Integrated Sensing and Communications}
\newacronym{mimo}{MIMO}{Multiple-Input Multiple-Output}
\newacronym[plural=FIDs]{fid}{FID}{Fisher Information Density}
\newacronym[plural=LSTMs]{lstm}{LSTM}{Long Short-Term Memory}
\newacronym{gdpr}{GDPR}{General Data Protection Regulation}
\newacronym{mno}{MNO}{Mobile Network Operator}
\newacronym{plr}{PLR}{Privacy Leak Ratio}

\usepackage{tcolorbox}
\tcbuselibrary{many}

\newtcbtheorem[]{alg}{Algorithm}{fonttitle=\bfseries}{alg}

\usepackage[norelsize,linesnumbered,ruled]{algorithm2e}
\SetKwRepeat{Do}{do}{while}
\makeatletter
\newcommand{\removelatexerror} {\let\@latex@error\@gobble}
\makeatother

\begin{document}
	
	\title{Balancing Functionality and GDPR-Driven Privacy in ISAC Trajectory Sharing}
	
	\author{
		\IEEEauthorblockN{Zexin~Fang\IEEEauthorrefmark{1},~Bin~Han\IEEEauthorrefmark{1},~Zhuojun~Tian\IEEEauthorrefmark{2} and~Hans~D.~Schotten\IEEEauthorrefmark{1}\IEEEauthorrefmark{3}}
		\IEEEauthorblockA{
  \IEEEauthorrefmark{1}{RPTU University Kaiserslautern-Landau, Germany}; \IEEEauthorrefmark{2}{KTH Royal Institute of Technology, Sweden}\\ 
  \IEEEauthorrefmark{3}{German Research Center for Artificial Intelligence (DFKI), Germany.}}
	}
	
	\bstctlcite{IEEEexample:BSTcontrol}
	
	\maketitle

	\begin{abstract}  
    \gls{isac} enables trajectory sharing 
that enhances beamforming, resource allocation, and cooperative perception, 
yet raises fundamental privacy concerns under the \gls{gdpr} data minimisation 
principle. This paper proposes a \gls{fid}-constrained 
trajectory sharing framework that enforces a local lower bound on estimation 
uncertainty, providing hard, quantifiable privacy guarantees by construction. 
Unlike fixed-noise approaches, the proposed method bounds the \gls{plr} regardless of sensing power or adversarial post-processing, 
ensuring that no trajectory segment can be reconstructed beyond a prescribed 
accuracy threshold. Simulations on the OpenTraj dataset demonstrate that the 
framework keeps the average \gls{plr} below $20$--$25\%$ and the maximum 
leakage segment duration under $2$--$2.5$ s, while preserving data utility 
for downstream tasks such as movement prediction. The resulting criterion is 
interpretable, model-agnostic, and compatible with \gls{gdpr}-compliant 
\gls{isac} system design.
		
	\end{abstract}
    
	
	\IEEEpeerreviewmaketitle
	
	\glsresetall

	\section{Introduction}\label{sec:introduction}
 \gls{isac} has recently emerged as a promising paradigm for next-generation 
wireless systems. By enabling trajectory storage and processing within a 
\gls{mno}, \gls{isac} sensing data can significantly enhance beamforming and 
radio resource allocation through movement prediction \cite{zhou2024temporal, yang2025cooperative, li2026attention, li2026recent}. Beyond network 
optimization, sharing such data can further contribute, in the longer term, 
to cooperative perception and collision avoidance for autonomous driving and robots \cite{liu2026goalo, ye2024isac,hu2025collaborative}. 
These capabilities are particularly valuable in dense urban environments, 
where high mobility and rapid channel dynamics pose fundamental challenges 
to conventional communication-only designs.
\begin{figure}[ht]
\centering
\includegraphics[width=0.95\linewidth]{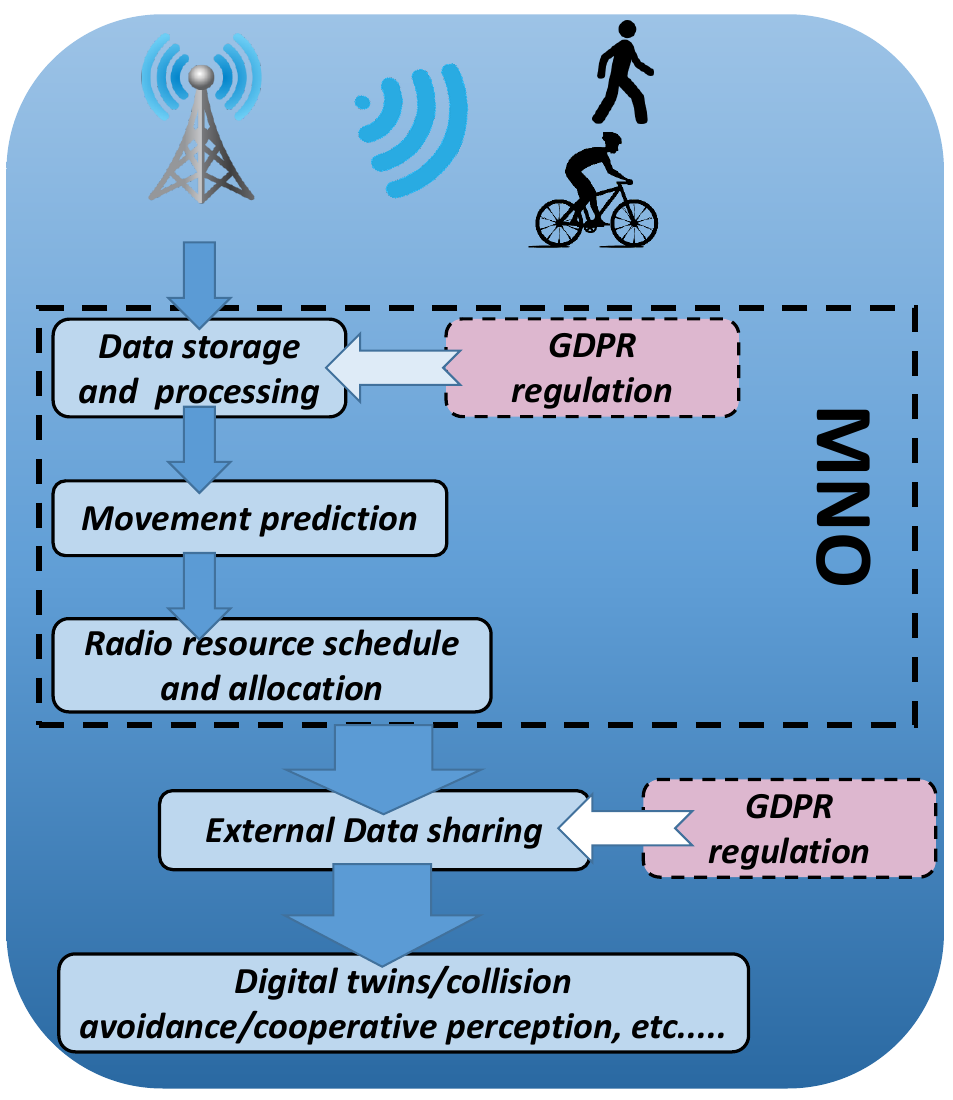}
\caption{\gls{isac} trajectory data flow within and beyond the \gls{mno}}
\label{fig:privacy_isac}
\end{figure}

At the same time, trajectory data handling in \gls{isac} systems raises 
fundamental data protection concerns under the EU \gls{gdpr}, illustrated in Fig. \ref{fig:privacy_isac}. In particular, 
the data minimisation principle mandates that personal data be limited to 
what is strictly necessary for a specified purpose, regardless of whether 
the data is shared with external parties or retained within trusted entities 
such as a \gls{mno}. However, the utility of trajectory data for movement prediction follows a straightforward principle: more data and higher data quality consistently 
yield better prediction accuracy. This also implies that sharing or 
processing such data carries significant privacy risks, as richer trajectory 
data inherently encodes more personal information. Privacy and utility are 
thus fundamentally at odds and cannot be jointly optimized without compromise. 
This tension motivates a clear design principle for a privacy-preserving 
framework: no more data than necessary should be shared once prediction 
accuracy is satisfactory, or equivalently, prediction accuracy should be 
maximized subject to a minimum privacy guarantee being met.

While data minimisation has been studied for static or 
application-layer data, how to formally enforce and evaluate it for 
trajectory data remains largely unexplored, particularly in the context 
of \gls{isac} systems. Unlike traditional radar systems, \gls{isac} sensing 
is inherently opportunistic and adaptive: the sensing update rate and accuracy 
are governed by radio resource allocation, mobility patterns, and sensing 
priorities, while complex urban propagation introduces time-varying and 
location-dependent sensing quality. These characteristics give rise to 
trajectory data with heterogeneous uncertainty and non-uniform sampling, 
posing unique challenges for \gls{gdpr}-compliant data minimisation. 
Crucially, purely quantitative assessments can be misleading: long, sparsely 
sampled, low-quality trajectories may be voluminous yet largely uninformative, 
whereas short, densely sampled, high-quality segments can enable precise 
localization and behavioral inference by malicious third parties.

To address this challenge, we propose a \gls{fid}-constrained trajectory 
sharing framework for \gls{isac} systems, using Fisher information as a 
principled proxy for privacy leakage. By enforcing a local lower bound on 
estimation uncertainty, the framework guarantees that no post-processing 
or denoising by an adversary can recover the original trajectory beyond a 
prescribed accuracy, fundamentally bounding privacy leakage at every segment 
of the shared data. This yields a quantitative and interpretable criterion 
for evaluating data minimisation that is agnostic to both sensing algorithms 
and adversarial models. Crucially, the enforced uncertainty bound translates 
directly into a worst-case privacy guarantee, making the proposed framework 
an explicit and principled enabler of \gls{gdpr}-compliant \gls{isac} system 
design.

The remainder of this paper is organized as follows. Sec.~\ref{sec:sys_model} presents the system model, including the Fisher Information-constrained trajectory sharing framework for \gls{mimo}-\gls{isac} systems and the privacy leak model, which together form the methodological foundation of our work. Sec.~\ref{sec:simu} evaluates the proposed approach through simulations, analyzing the trade-off between privacy preservation and data utility using complementary metrics for both aspects. Finally, Sec.~\ref{sec:conclu} concludes the paper and discusses future research directions.

\section{System model and methodology}\label{sec:sys_model}
% \begin{figure}[t]
% 	\centering
% 	\includegraphics[width=0.82\linewidth]{icasslot_cropped.pdf}
% 	\caption{Illustration of measurement denoising window for ultra-short TDOA sequences}
% 	\label{fig:denoise_proccess}
% \end{figure}
\subsection{Sensing model for Smart and multi-functional \gls{isac} system}
We consider a dual-task \gls{mimo}-ISAC system, where a \gls{bs} equipped with $N_t$ transmit and $N_s$ receive antennas simultaneously communicates with a multi-antenna communication user and senses multiple targets of interest. The system can dynamically switch between operational modes depending on the spatial distribution of the targets \cite{fanliu2020isac, liu2025isacmode, yin2025target,zhao2025target}.

The allocation of sensing symbols and transmit power can be dynamically 
adjusted according to channel conditions or target priority. Within a given 
sensing-and-communication slot, a high-priority target may be sensed multiple 
times to improve temporal resolution, while lower-priority targets are sensed 
less frequently. This allows the system to adaptively increase the sensing 
frequency for critical targets, ensuring finer tracking performance while 
respecting the overall resource budget.

Modern \gls{isac} architectures are further expected to support 
sensing-performance-aware resource management. Based on the current channel 
state information, beamforming configuration, and sensing parameters, the 
system can estimate the achievable sensing accuracy, typically characterized 
through the \gls{crb}~\cite{liuzhu2026isac, crb2024ren}. Such estimates 
enable quality-aware resource allocation by adapting sensing power, symbol 
allocation, and scheduling decisions to meet desired performance targets. 
It should be noted that most current \gls{isac} systems operate in the 
sub-$10$ GHz range, where the Doppler shift of slow-moving targets is 
negligible and can therefore be ignored. Under this assumption, the 
\gls{crb} for sensing target $i$ is given by \cite{liuzhu2026isac}
\begin{equation}
\mathrm{CRB}_i = \frac{1}{\beta_i \, m_{s,i} \, \mathbf{a}_i^H \mathbf{Q}_{s,i} \mathbf{a}_i},
\label{eq:crb}
\end{equation}
where
\begin{itemize}
    \item $\beta_i$ represents the propagation and target-dependent gain, 
    encompassing path loss and radar cross-section;
    \item $m_{s,i}$ denotes the number of sensing symbols allocated to 
    target $i$ within one sensing update;
    \item $\mathbf{a}_i$ is the steering vector corresponding to the 
    target direction;
    \item $\mathbf{Q}_{s,i}$ is the transmit covariance matrix for sensing, 
    whose trace reflects the total sensing transmit power.
\end{itemize}

Building on this capability, we consider {Fisher information--constrained 
trajectory sharing}, where the amount of trajectory information disclosed across 
network entities is regulated according to the estimated sensing accuracy. In 
this framework, the sensing Fisher information serves as a principled proxy for 
the informativeness of the sensed trajectory, enabling the system to balance 
privacy protection against the need to retain sufficient data quality for 
downstream tasks. Since Fisher information is simply the inverse of the 
\gls{crb}, it follows directly from~\eqref{eq:crb} that
\begin{equation}
\mathcal{I}_i = \beta_i \, m_{s,i} \, \mathbf{a}_i^H \mathbf{Q}_{s,i} \mathbf{a}_i.
\label{eq:fi}
\end{equation}
In a sequential sensing and estimation framework, each observation of a 
target contributes to reducing the uncertainty of its estimated trajectory. 
Let $\mathcal{I}_i(k)$ denote the Fisher information obtained from the 
$k$-th sensing update of target $i$, occurring at time $t_k$. The total 
accumulated Fisher information over a time window $\Delta T$ containing 
$K$ updates is
\begin{equation}
\mathcal{I}_i^{\text{sum}}(K) = \sum_{k=1}^{K} \mathcal{I}_i(k),
\end{equation}
and the average Fisher information rate over $\Delta T$ is
\begin{equation}
\bar{\mathcal{I}}_i = \frac{1}{\Delta T}\sum_{k=1}^{K} \mathcal{I}_i(k).
\end{equation}
As the update interval shrinks, i.e., $\Delta T / K \to 0$, this average 
rate converges to a continuous-time density, which we define as the 
\gls{fid}:
\begin{equation}
\mathcal{J}_i(t) = \lim_{\Delta T / K \to 0} 
\frac{1}{\Delta T}\sum_{k=1}^{K} \mathcal{I}_i(k).
\end{equation}
In practice, since sensing updates occur at discrete instants, the 
\gls{fid} is evaluated in piecewise form as
\begin{equation}
\mathcal{J}_i(t) = \frac{\mathcal{I}_i(k)}{t_k - t_{k-1}}, 
\quad t \in (t_{k-1}, t_k],
\end{equation}
By constraining $\mathcal{J}_i(t)$, the privacy 
of every segment is preserved by construction: no signal processing or 
denoising applied to any segment of the shared trajectory can reduce the 
reconstruction error below a guaranteed minimum threshold.

\subsection{Privacy leak model for trajectory sharing}

Let $\tilde{\bm{m}}_i(t_k)$ denote the trajectory data of target $i$ shared 
by the \gls{isac} system, $\overline{\bm{m}}_i(t_k)$ its reconstruction from 
the shared data, and $\bm{m}_i(t_k)$ the ground-truth trajectory, where each 
sample comprises the coordinates of the target relative to the base station, 
i.e., $[x_{t_k}, y_{t_k}]^T$, derived from range and angular measurements. 
To simplify the analysis, we adopt a tractable angular error model 
parameterized by the \gls{snr}; the detailed error modelling is provided 
in~\cite{sensing2025fang}. The point-wise reconstruction error is defined as
\begin{equation}
e_i(t_k) = \| \overline{\bm{m}}_i(t_k) - \bm{m}_i(t_k) \|.
\end{equation}

We define the Privacy Leak Ratio (PLR) as the fraction of trajectory points 
reconstructable by an adversary:
\begin{equation}
\text{PLR}_i = 
\frac{
    \left| \{ t_k \in \mathcal{T}_i : e_i(t_k) \le \epsilon \} \right|
}{
    \left| \mathcal{T}_i \right|
},
\end{equation}
where $\mathcal{T}_i$ denotes the set of sampling instants of the trajectory, 
$e_i(t_k)$ is the reconstruction error at time $t_k$, and $\epsilon$ is the 
minimum error threshold required to guarantee privacy. A lower PLR indicates 
stronger privacy protection.

To monitor PLR, \gls{fid} can be controlled while a \gls{isac} system sharing data by adding controlled error $\Delta e_i(t_k)$,
\begin{equation}
\Delta e_i(t_k) \sim  \mathcal{N}(0,(\Delta \sigma(t_k))^2), 
\end{equation}
\begin{equation}
\Delta \sigma(t_k) =
\begin{cases}
0, & \mathcal{J}_i(t) \le \eta , \\
\alpha\left(\beta - e^{-(\frac{\mathcal{J}_i(t)}{\eta}-1)}\right), & \mathcal{J}_i(t) > \eta .
\end{cases}\label{eq:error_con}
\end{equation}
\begin{equation}
\tilde{\bm{m}}_i(t_k) =  \hat{\bm{m}}_i(t_k) + \Delta e_i(t_k), 
\end{equation}
where $\hat{\bm{m}}_i(t_k)$ denotes the raw \gls{isac} measurements without any processing. In Eq.~\eqref{eq:error_con}, $\Delta \sigma(t_k)$ is a thresholded, saturating function of the \gls{fid}, adding no noise for low-Fisher-information segments and smoothly increasing to a maximum for high-Fisher-information segments, thus balancing the overall privacy–utility tradeoff.

\begin{figure}[t]
    \centering
    \begin{subfigure}{.91\linewidth}
        \centering     
        \includegraphics[width=\linewidth]{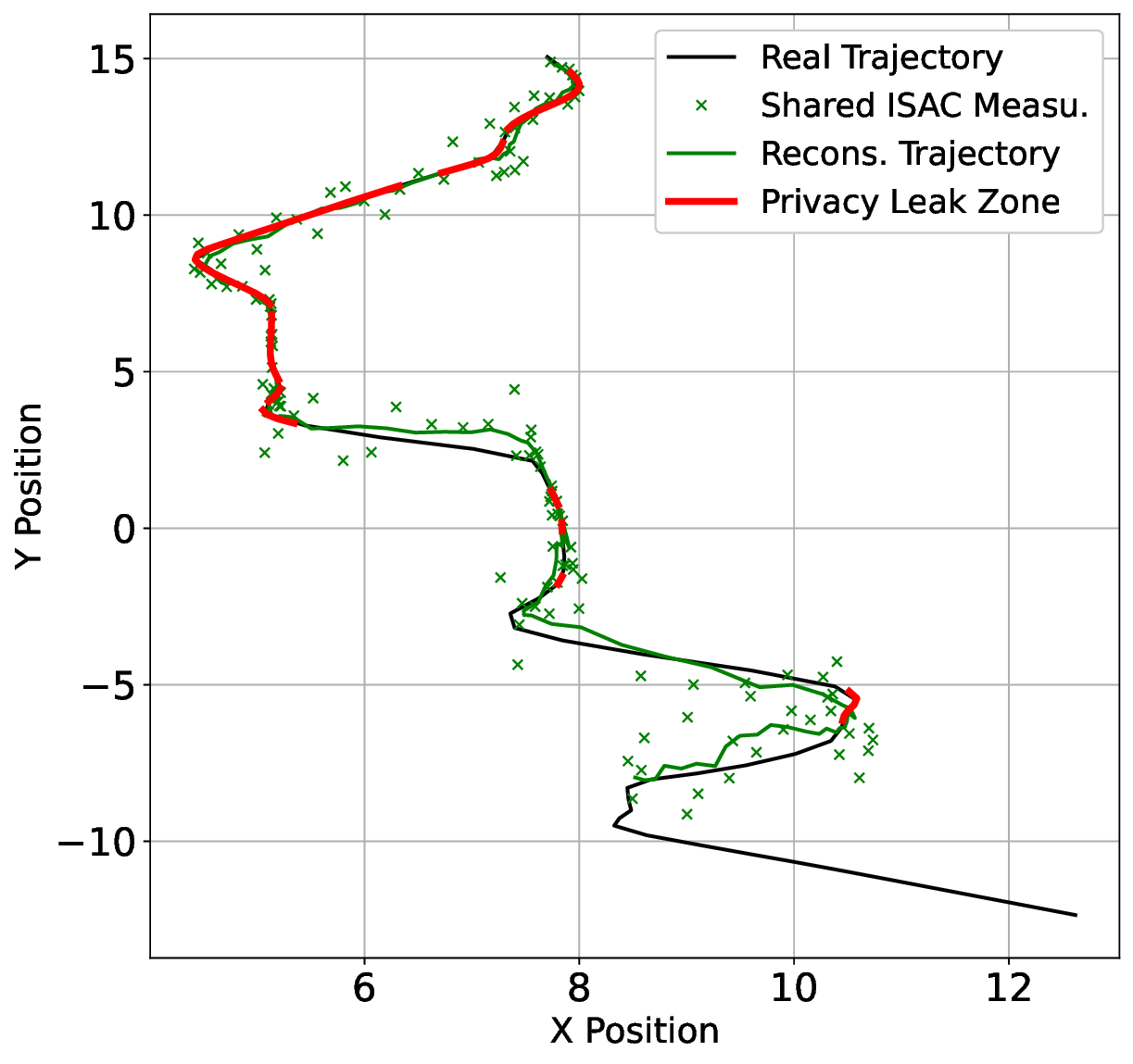}
        \subcaption{Privacy leakage without \gls{fid}-constrained sharing}
        \label{subfig:lowalti}
    \end{subfigure}
    \begin{subfigure}{.91\linewidth}
        \centering
        \includegraphics[width=\linewidth]{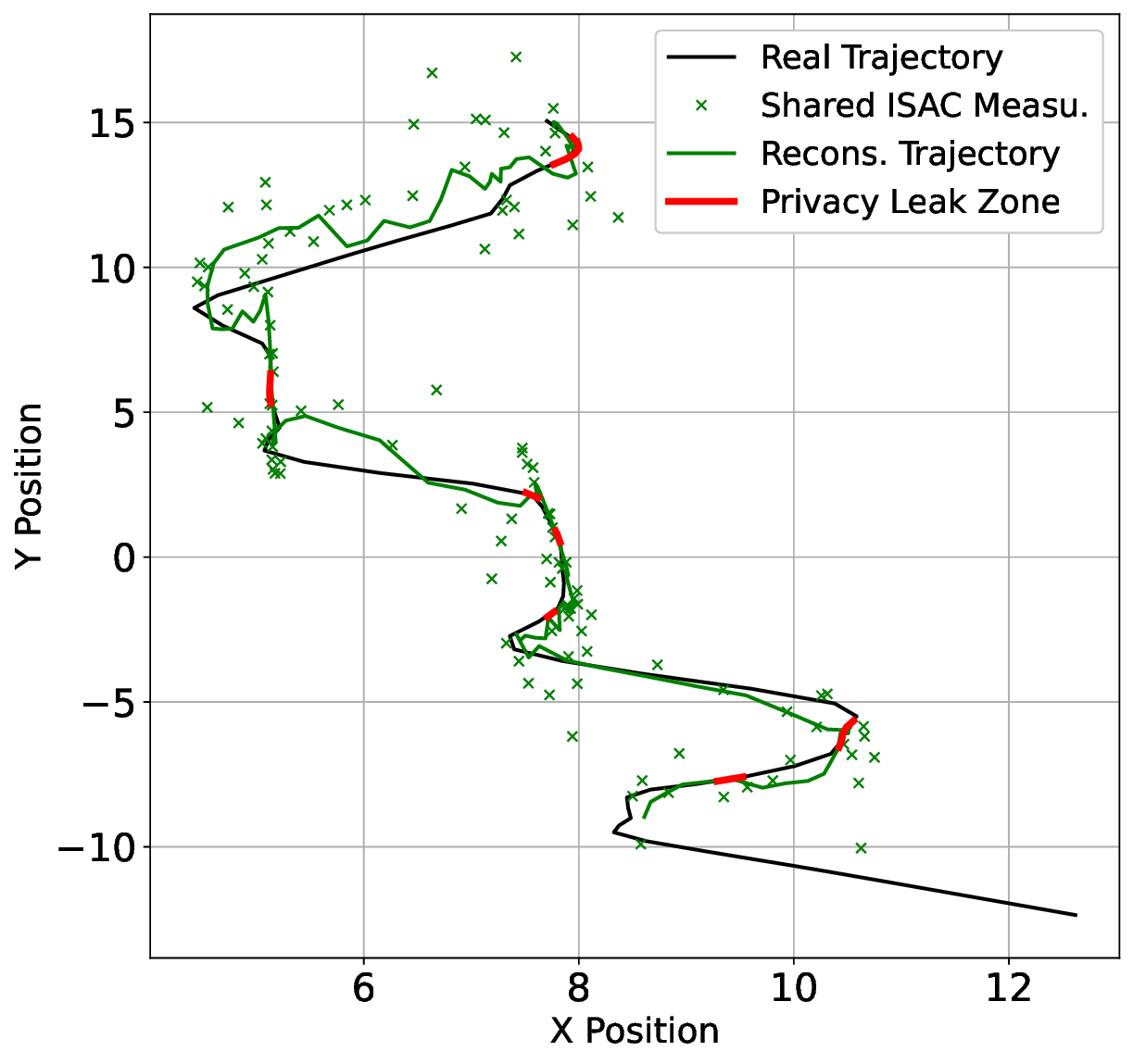}
        \subcaption{Privacy leakage with \gls{fid}-constrained sharing}
        \label{subfig:normal}
    \end{subfigure}
    \caption{Demonstration of trajectory privacy leakage. The reconstructed 
    trajectory is obtained from heavily smoothed \gls{isac} measurements 
    over 7 points.}
    \label{fig:privacyleak}
\end{figure}

We demonstrate the privacy protection performance in Fig.~\ref{fig:privacyleak}. 
The trajectory data are drawn from the OpenTraj dataset~\cite{OpentrajJavad}, 
a widely used benchmark comprising real-world pedestrian trajectories collected 
across diverse public scenes, including university campuses, plazas, and 
intersections. A subset of trajectories is selected to emulate user mobility 
within the service area of a single base station positioned at $[5, 30]$, 
with the coordinate system normalized to match the simulation environment. 
The remaining simulation parameters follow those listed in 
Tab.~\ref{tab:params}. As evident from the comparison, constraining the 
\gls{fid} of the shared data substantially reduces the PLR. Despite heavy 
smoothing of the reconstructed trajectory, the deviation from the original 
exceeds $1$ meter in many cases, posing a significant obstacle to behavioral 
inference by an adversary, particularly in intersections and streets.
\begin{figure*}[!ht]
    \centering
   \begin{subfigure}[b]{0.329\textwidth}
        \centering
        \includegraphics[width=\textwidth]{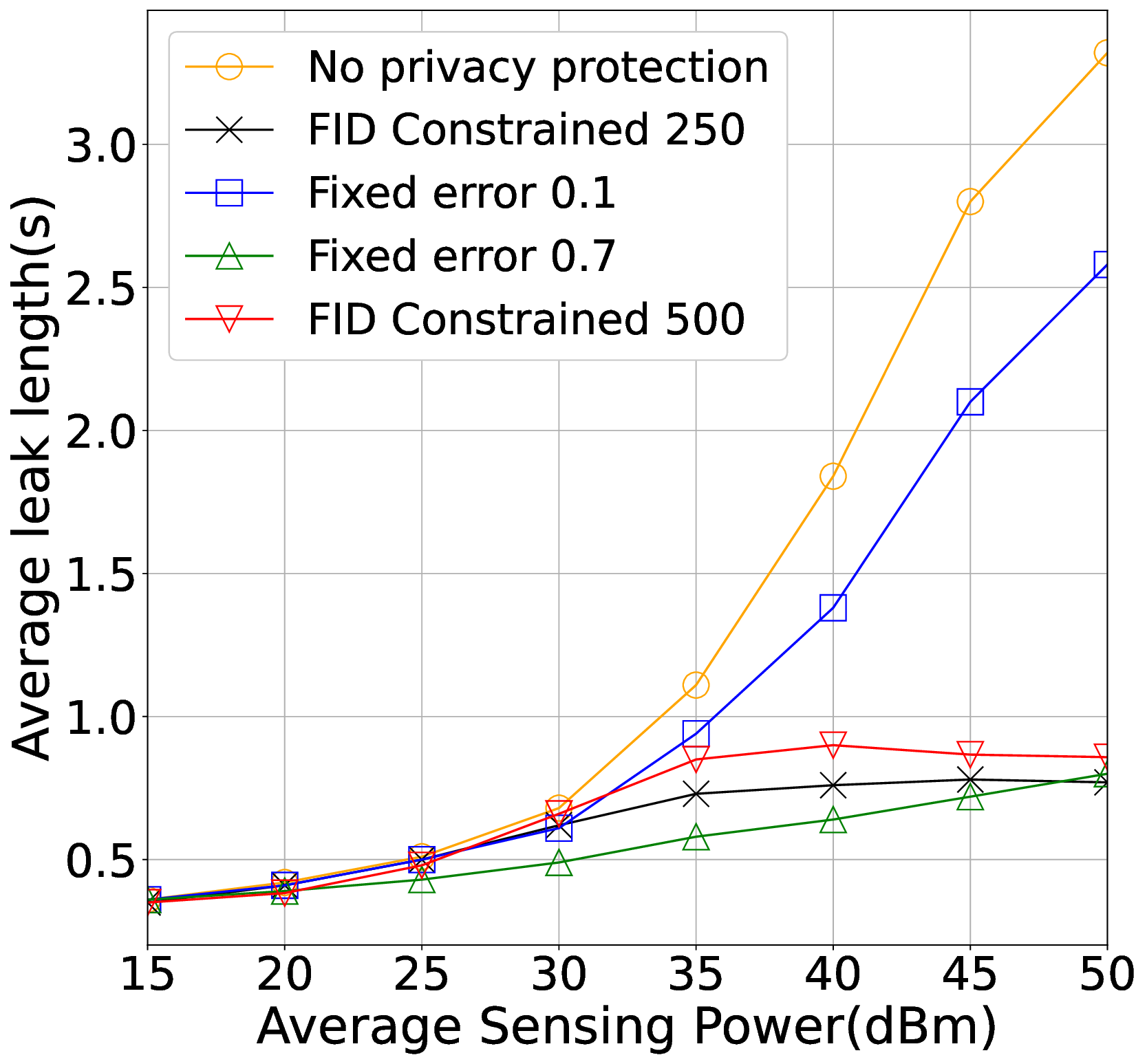}
        \caption{}
    \end{subfigure}
   \begin{subfigure}[b]{0.329\textwidth}
        \centering
        \includegraphics[width=\textwidth]{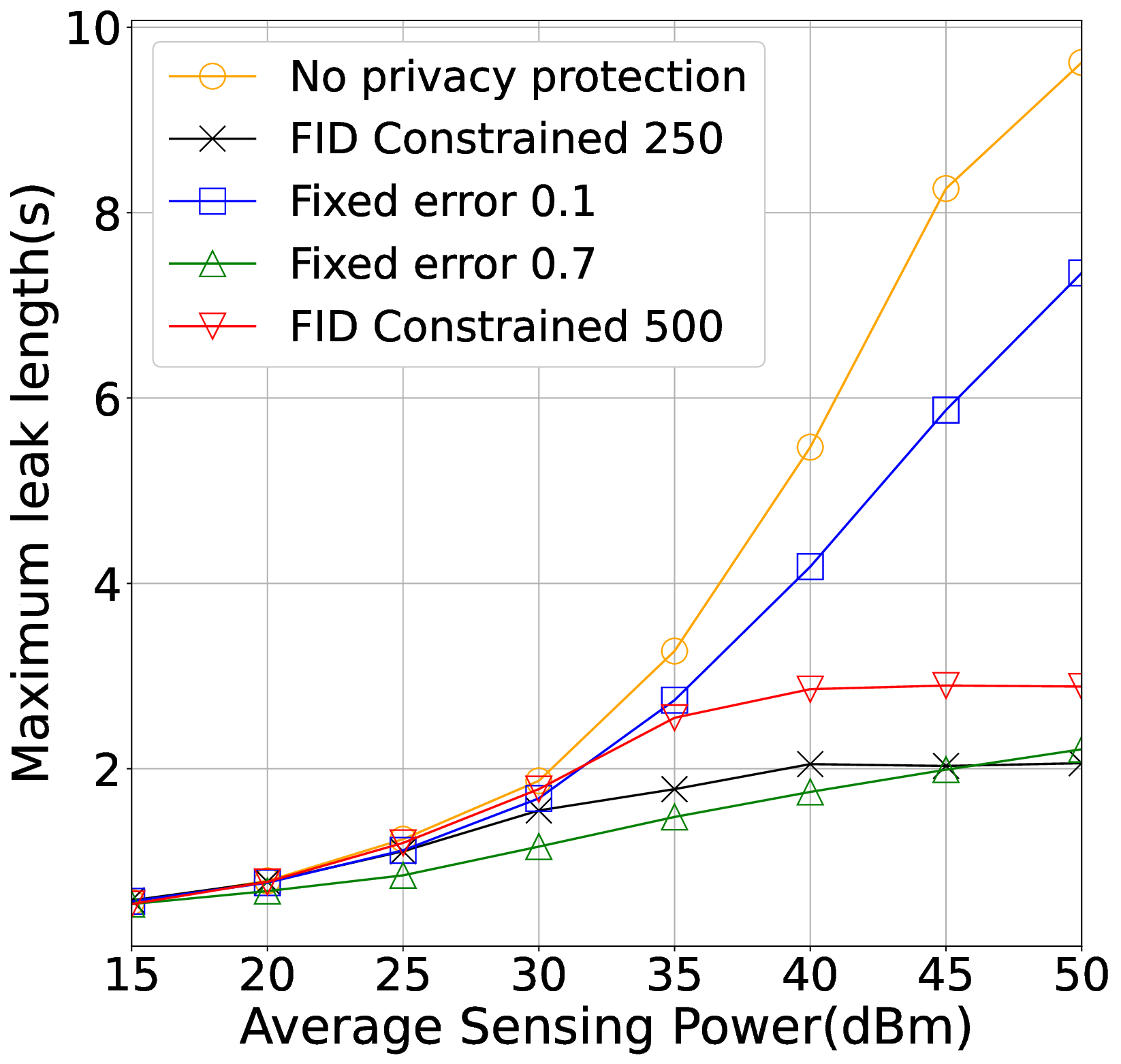}
        \caption{}
    \end{subfigure}
   \begin{subfigure}[b]{0.329\textwidth}
        \centering
        \includegraphics[width=\textwidth]{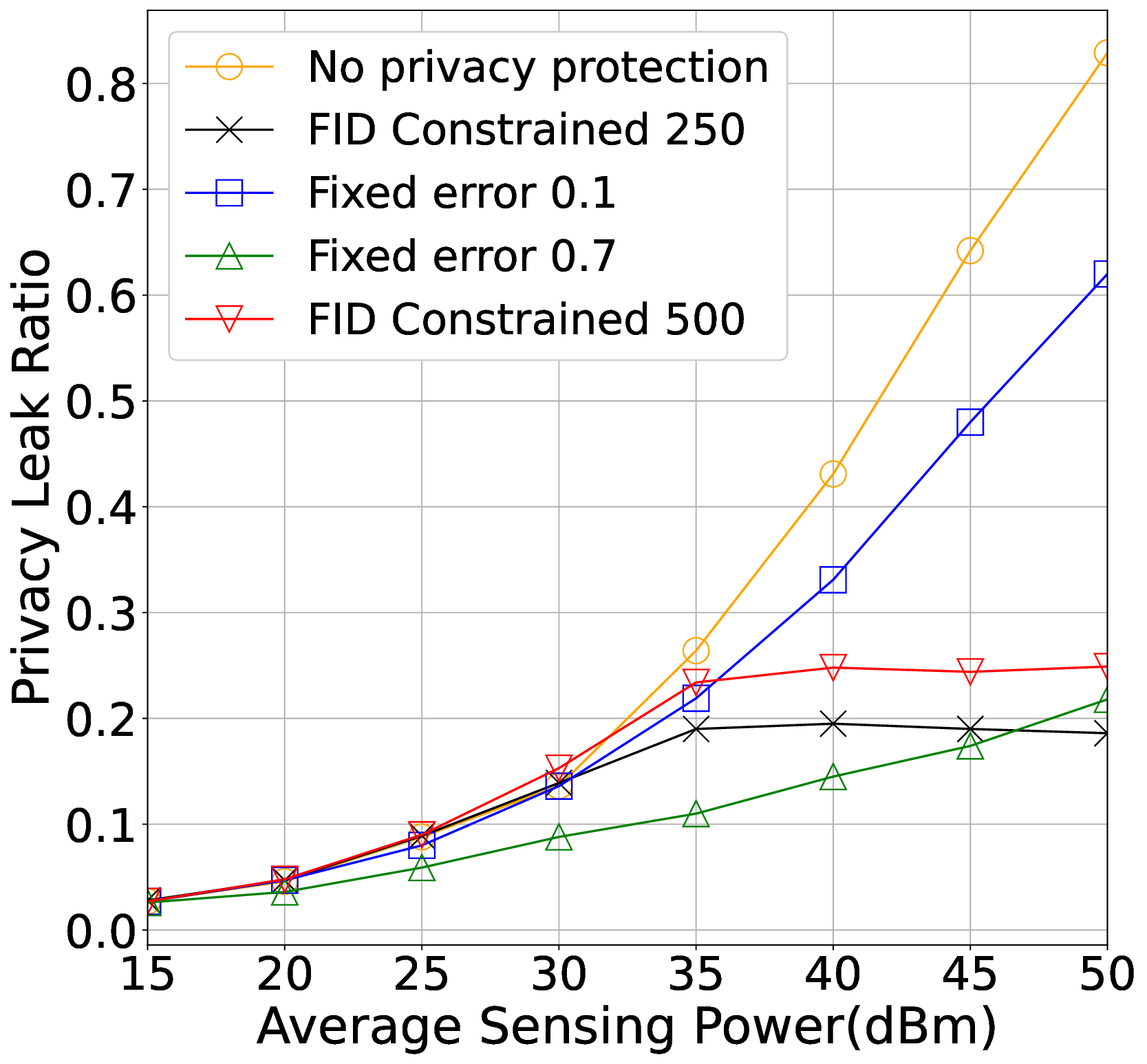}  
        \caption{}
    \end{subfigure}\caption{Privacy evaluation with different metrics: (a) average
privacy leak segment duration; (b) the maximum privacy leak
segment duration; (c) average privacy leak ratio of the entire trajectory.}\label{fig:ls_Pri}, 
    \end{figure*}
 \begin{figure*}[!ht]
    \centering
   \begin{subfigure}[b]{0.329\textwidth}
        \centering
        \includegraphics[width=\textwidth]{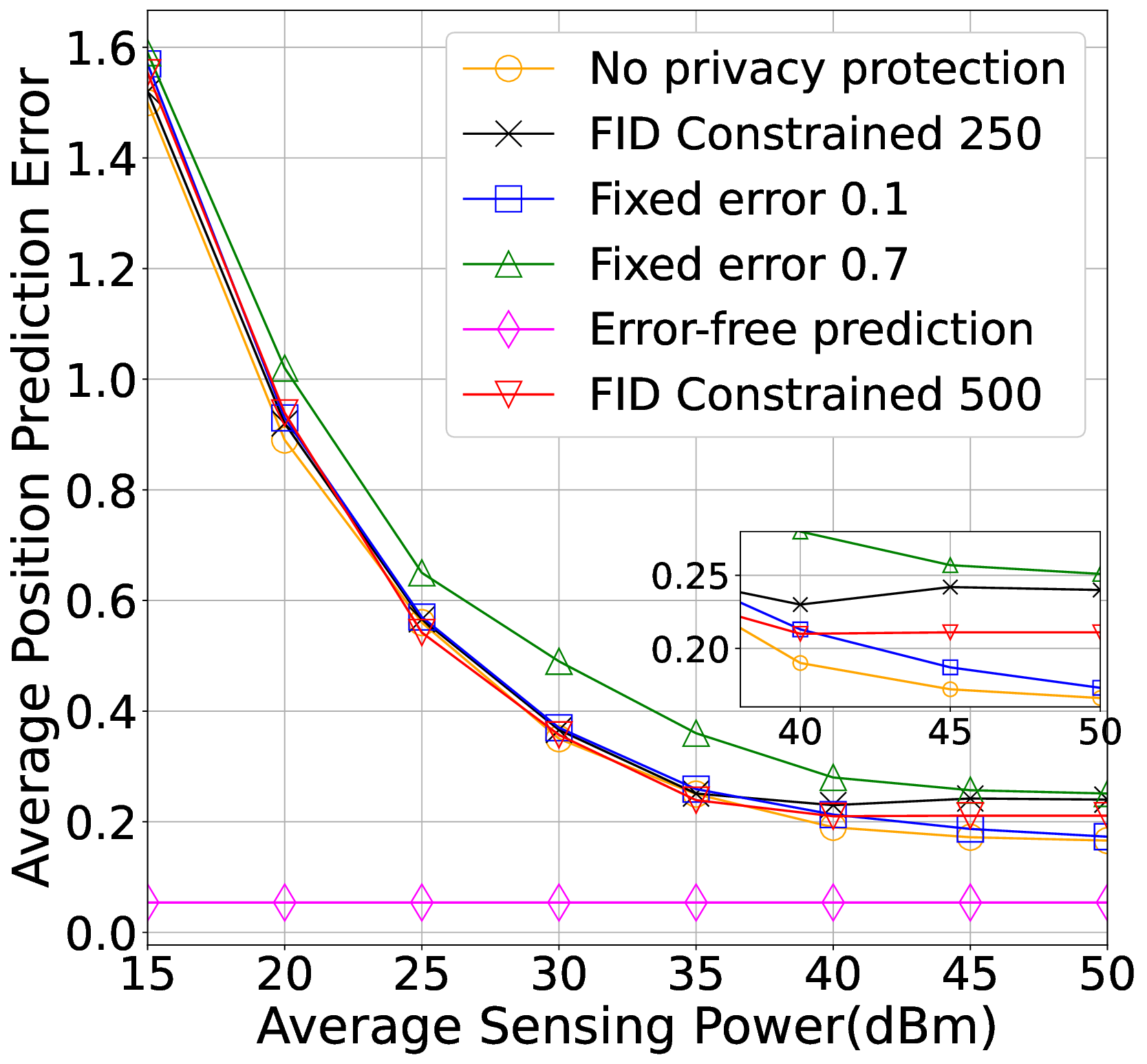}
        \caption{}\label{fig:bsglo}
    \end{subfigure}
   \begin{subfigure}[b]{0.329\textwidth}
        \centering
        \includegraphics[width=\textwidth]{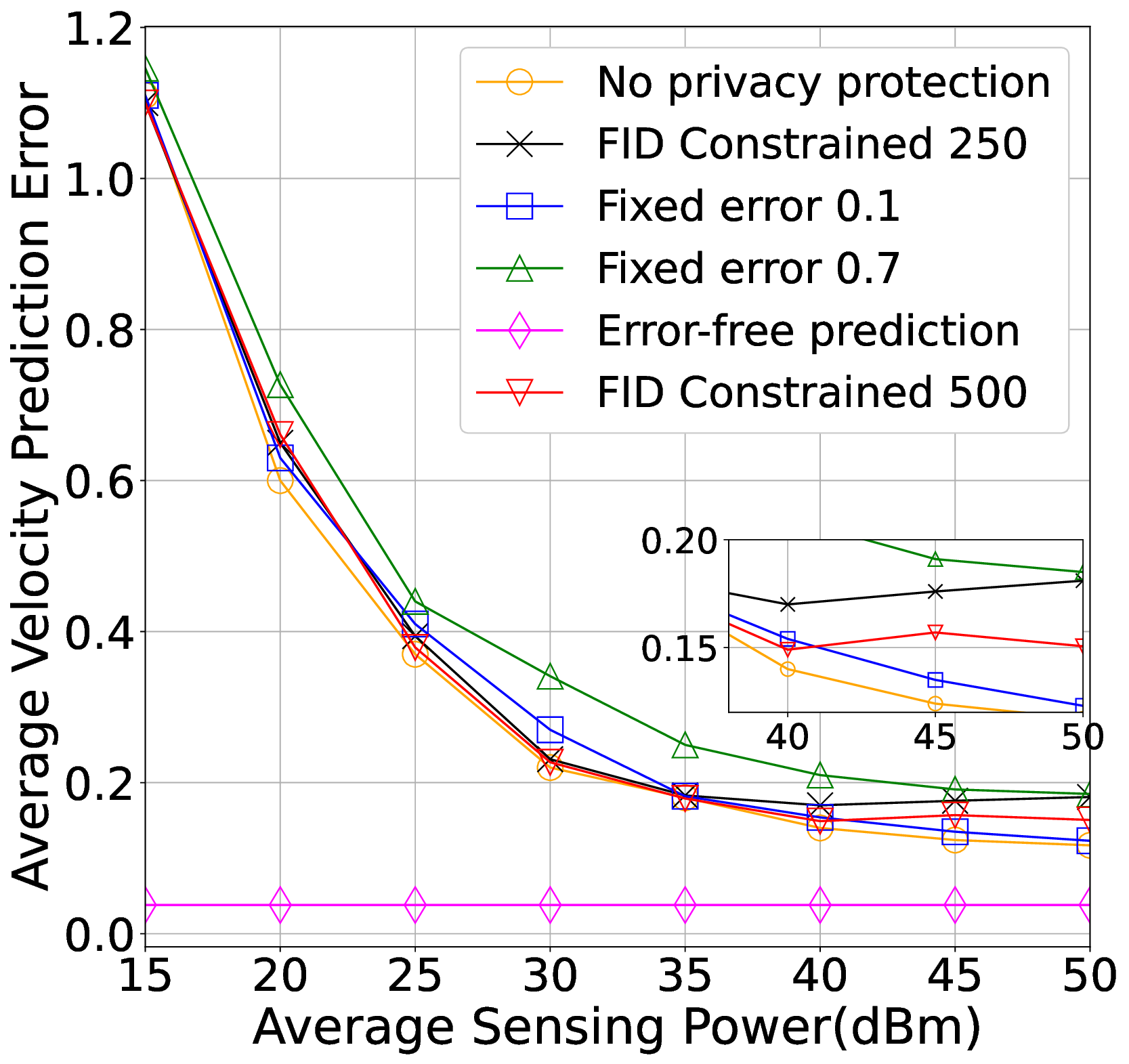}
        \caption{}\label{fig:bssel}
    \end{subfigure}
   \begin{subfigure}[b]{0.329\textwidth}
        \centering
        \includegraphics[width=\textwidth]{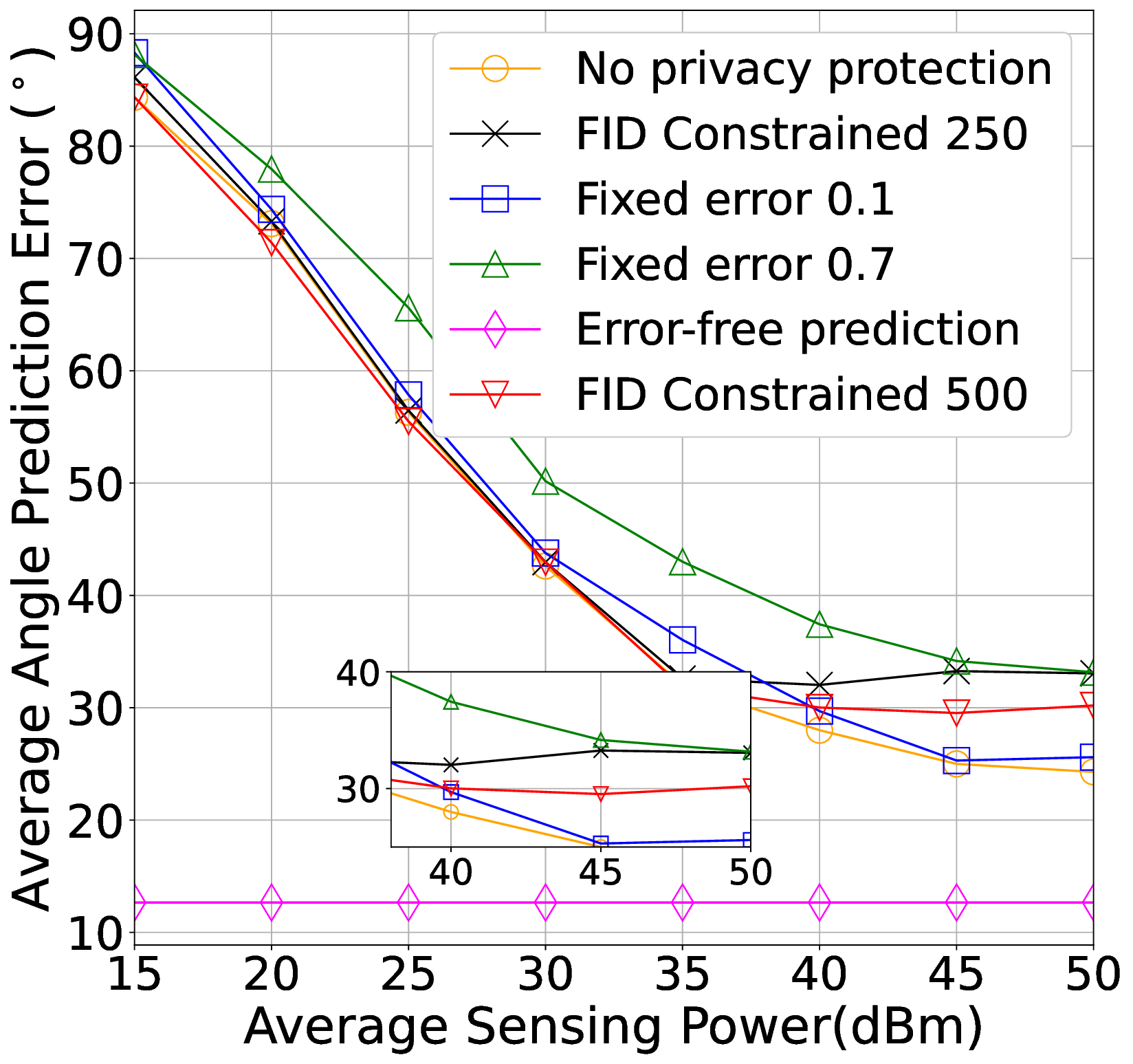}  
        \caption{}\label{fig:bsfoc}
    \end{subfigure}\caption{Utility evaluation: (a) one-second-ahead position error; (b) absolute velocity error; (c) heading angle error. The error-free baseline is obtained 
by evaluating the model on the original trajectory data without any sensing 
error. As a simple \gls{lstm} model is employed and pedestrian trajectories 
are inherently complex, the heading angle prediction exhibits comparatively 
larger errors.}\label{fig:ls_ul}
    \end{figure*}
\section{Simulation and Evaluation}\label{sec:simu}
Having demonstrated the privacy protection enabled by \gls{fid}-constrained 
trajectory sharing, we now evaluate the trade-off between privacy preservation 
and data utility. As observed in Fig.~\ref{fig:privacyleak}, privacy leakage 
may occur in continuous segments of the trajectory, where longer leakage 
segments pose a greater risk of privacy exposure. For \textit{privacy 
evaluation}, we adopt three complementary metrics: the average leakage 
segment duration, the maximum leakage segment duration, and the average 
PLR over the entire trajectory. For \textit{utility evaluation}, we assess 
the effectiveness of the shared trajectory data for movement prediction 
using a lightweight \gls{lstm} model trained on $10{,}000$ trajectory 
segments of $12$ s duration. Prediction performance is evaluated by feeding 
\gls{fid}-constrained trajectory segments into the trained model, measured 
across three metrics: one-second-ahead position error, absolute velocity 
error, and heading angle error.

The base station is fixed at $[5, 30]$ in the normalized coordinate system. 
To emulate sensing resource reallocation in practical systems, the update 
rate refreshes every $5$ seconds and is drawn uniformly from $[2, 4]$ 
samples/s, averaging at $3$ samples/s. Channel conditions switch between 
\gls{los} and \gls{nlos} following a probabilistic blockage model, reflecting 
the dynamic propagation environment typical of urban settings. Trajectory 
data are drawn from the OpenTraj dataset~\cite{OpentrajJavad}, spanning 
durations of $10$ to $100$ seconds and encompassing a wide variety of 
mobility patterns. Fixed measurement errors of $\Delta\sigma_i = 0.1$ and 
$\Delta\sigma_i = 0.7$ are used as baselines for comparison, while multiple 
\gls{fid} thresholds are evaluated to demonstrate varying levels of privacy 
requirement and their associated trade-offs with data utility. The remaining simulation parameters are summarized in Tab.~\ref{tab:params}. 
All simulation results are averaged over Monte Carlo simulations conducted 
across diverse trajectory.
\begin{table}[]
		\centering
		\caption{Simulation setup 1}
		\label{tab:params}
		\begin{tabular}{>{}m{0.1cm} | m{1.0cm} l m{4.0cm}}
			\toprule[2px]
%			\rowcolor{white}
			&\textbf{Parameter}&\textbf{Value}&\textbf{Remark}\\
			\midrule[1px]        
			&$f_c$&$3.5$ Ghz& Carrier frequency\\

			&$K$&$(0.1,3.0)$& Rician factors\\
			&$N_p$&$4$& Number of multipath\\
            & $\tau_\text{max}$ & 2e-7 s &  Maximum delay spread\\ 
            & $\overline{n}$ & 2.7 &  Average path loss component\\ 
            & $N_\text{o}$ & -91 dBm &  Noise floor\\ 
            & $\beta$ & 100 Mhz &  Bandwidth\\ 
            & $P_{\text{tx}}$ & $15$--$50$ dBm & Average effective transmit power \\
            &$g_b$ & $\sim \mathcal{N}(0,2)$ dB & Beam misalignment-induced fluctuation \\
            \midrule[1px]
            \multirow{-14.0}{*}{\rotatebox{90}{\textbf{Sensing \& Channel}}}
            & $\epsilon$&$0.3$ m& Minimum error threshold\\ 
            & $\eta$&$50, 250$& \gls{fid} threshold\\ 
			\multirow{-2.7}{*}{\rotatebox{90}{\textbf{Privacy Eva.}}}&$\alpha$&$0.5$ & Noise perturbation parameter in Eq.~(\ref{eq:error_con}) \\
            &$\beta$&$1.5$ & Noise perturbation parameter in Eq.~(\ref{eq:error_con})\\
            \midrule[1px] 
            & $d_{\text{in}},d_{\text{out}}$ & $2$ & Input and output feature dimension (2D trajectory coordinates$\rightarrow$2D position prediction) \\
            & $H$ & $64$ & Hidden state dimension of LSTM \\
            & $L$ & $3$ & Number of LSTM layers \\
            & $\eta$ & $1.6\times10^{-3}$ & Learning rate of Adam optimizer \\
            \multirow{-6.7}{*}{\rotatebox{90}{\textbf{Utility Evaluation}}}& $B$ & $64$ & Batch size \\
            & $N_{\text{epoch}}$ & $1000$ & Number of training epochs \\
        \bottomrule[2px]
		\end{tabular}
	\end{table}

Fig.~\ref{fig:ls_Pri} examines the privacy protection performance as a 
function of average effective sensing power, reflecting the dynamic sensing 
resource allocation of the \gls{isac} system. As expected, privacy leakage 
risk grows with increasing sensing power across all schemes. Adding a fixed 
error of $\Delta\sigma_i = 0.1$ offers negligible protection, while 
$\Delta\sigma_i = 0.7$ yields a more noticeable effect, yet may still fail 
under sufficiently high sensing power. The proposed method, by contrast, 
maintains consistently bounded privacy leakage regardless of sensing power: 
the average PLR and maximum leakage segment duration are reliably kept below 
$20\%$ or $25\%$, and $2$ s or $2.5$ s, respectively, depending on the 
selected \gls{fid} threshold. This hard-guarantee behavior fundamentally 
distinguishes the proposed approach from fixed-noise baselines and makes it 
inherently more compatible with regulatory privacy frameworks.

The utility evaluation in Fig.~\ref{fig:ls_ul} further highlights the advantage of the proposed method. 
In the $30$--$40$ dBm range, $\Delta\sigma_i = 0.7$ introduces a significant 
performance gap relative to other approaches, as the excessive perturbation 
severely degrades data utility. The proposed method, by contrast, barely 
impacts utility when the average sensing power is below $30$ dBm. Beyond this 
point, as the minimum privacy requirement becomes binding, utility is 
correspondingly lower-bounded, a deliberate and desirable trade-off that 
aligns with the data minimisation principle outlined in \gls{gdpr}: only the 
minimum necessary data quality is exposed, and no more.

\section{Conclusion and outlook}\label{sec:conclu}
This paper proposed a \gls{fid}-constrained framework for \gls{gdpr}-compliant 
trajectory sharing in \gls{isac} systems. By enforcing a local lower bound 
on estimation uncertainty via the \gls{crb}, the framework 
guarantees that trajectory reconstruction error cannot fall below a prescribed 
threshold regardless of sensing power or adversarial post-processing, a 
property unattainable with fixed-noise approaches. Results confirm bounded 
privacy leakage across heterogeneous mobility patterns while preserving 
utility for downstream tasks.

Future work will formalize the privacy-utility trade-off as a Pareto 
optimization problem, derive closed-form expressions under varying channel 
and resource conditions, and adversarial 
inference models to further strengthen the framework.

          \section*{Acknowledgment}
	This work is supported by the Federal Ministry of Research, Technology and Space of Germany via the project
Open6GHub+ (16KIS2406). B. Han (bin.han@rptu.de) is the
corresponding author. 
    
\bibliographystyle{IEEEtran}
\bibliography{references}

\end{document}